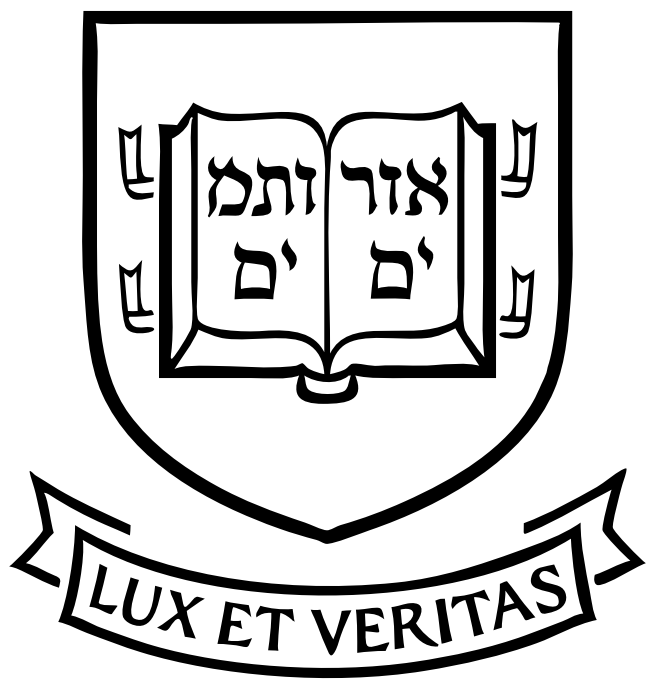

# Yale University
# Department of Computer Science

**Mosaic: Policy Homomorphic Network Extension**

L. Erran Li    M. F. Nowlan    C. Tian    Y. R. Yang    M. Zhang
Bell Labs    HUST    Microsoft Research    Yale

YALEU/DCS/TR-1427
February 2010

# Mosaic: Policy Homomorphic Network Extension

L. Erran Li    M. F. Nowlan    C. Tian    Y. R. Yang    M. Zhang
Bell Labs    HUST    Microsoft Research    Yale

**Abstract**

With the advent of large-scale cloud computing infrastructure, network extension and migration has emerged as a major challenge in the management of modern enterprise networks. Many enterprises are considering extending or relocating their network components, in whole or in part, to remote, private and public data centers, in order to attain scalability, failure resilience, and cost savings for their network applications. In this paper, we conduct a first rigorous study on the extension and migration of an enterprise network while preserving its performance and security requirements, such as layer 2/layer 3 reachability, and middle-box traversal through load balancer, intrusion detection and ACLs. We formulate this increasingly important problem, present preliminary designs, and conduct experiments to validate the feasibility of our designs.

## 1 Introduction

Due to enterprise dynamics (*e.g.*, expansion into a new site), hardware consolidation, and the emergence of cloud computing infrastructures, network extension and migration has become a major challenge in the management of modern enterprise networks. On the one hand, as many enterprises run out of space in their existing data centers [7], they need to extend or relocate their network to new private data centers. On the other hand, recent emergence of public cloud computing infrastructure provides enormous opportunities for an enterprise to either replace or complement its existing servers with computing resources in the cloud, in order to take advantage of improved efficiency and reliability. We refer to the private or public data centers that an enterprise extends to as the *remote data centers*.

Despite their potential business benefits and needs, such extension and migration can become quite complex and pose substantial challenges to the operation of enterprise network infrastructure. In particular, such extensions often have to be incremental instead of a complete restructuring of the existing network infrastructure. Thus, a seemingly small extension can be extremely challenging to handle in practice.

Consider a simple example of relocating a set of application servers from one data center of the enterprise to a remote data center (*e.g.*, another private or public cloud data center). These servers usually have complex communication patterns regulated by network policies such as traversal of firewalls and intrusion detection systems before being reached. Furthermore, an enterprise network may enforce network policies using a variety of techniques including routing design, topology design, and deployment of policy boxes at strategic locations. Some such techniques, such as deployment at topology cuts, can be implicit without any explicit representation. Consequently, it can be extremely challenging to take these servers out of their current "context" and place them into another "context" while preserving existing network policies. Manual reconfiguration, although maybe feasible for small networks, can no longer satisfy the need to scalable to large systems.

There are two common ways to connect an enterprise network to a remote data center. In one extreme, a remote data center may belong to the same enterprise, allowing plenty of flexibility in constructing network topology and policy boxes inside the remote data center. In the other extreme, a remote data center may belong to a public cloud provider, imposing substantial restrictions on the connection and layout of the remote data center.

We present Mosaic, a first framework for network extension and migration while preserving enterprise network policies. Mosaic introduces two key notions — way-points and scopes — to capture network policy constraints during network extension. Moreover, Mosaic includes two simple and yet powerful primitives named proxy and mirror to implement network extensions with provable guarantees. Guided by the policy contraints and utilizing the primitives, a Mosaic extension algorithm computes efficient network extension strategy. We refer to policy-preserving network extension as *policy homomorphic network extension*.

We proceed by presenting a rigorous analysis of the requirements and constraints of preserving policies during migration. We then evaluate our novel network extension algorithm in a large campus network setting. Our preliminary results indicate that Mosaic extension algorithm performs far better than a naive server relocation algorithm in terms of number of policy violations.

## 2  Motivating Example

We start with a motivating example for the consolidation of resources into either a private data center or a public data center such as Amazon's EC2. Figure 1 shows part of a real network. The network is a relatively standard three-tiered design, hosting multiple applications of the organization. The figure removes the exact model numbers of the devices. For reliability, each logical network device (*e.g.*, firewall $F_i$, load balancer $LB_i$, intrusion prevention system $IPS_i$, and switch $S_j$, $i = 1,2$, $j = 1, \cdots, 5$) represents two identical physical devices. One is active, and the other standby. To make the figure easy to read, we draw only one such device. Note that $LB_1, LB_2$ and *CE* are layer 3 (L3) devices; servers are endpoints; and the rest are layer 2 (L2) devices.

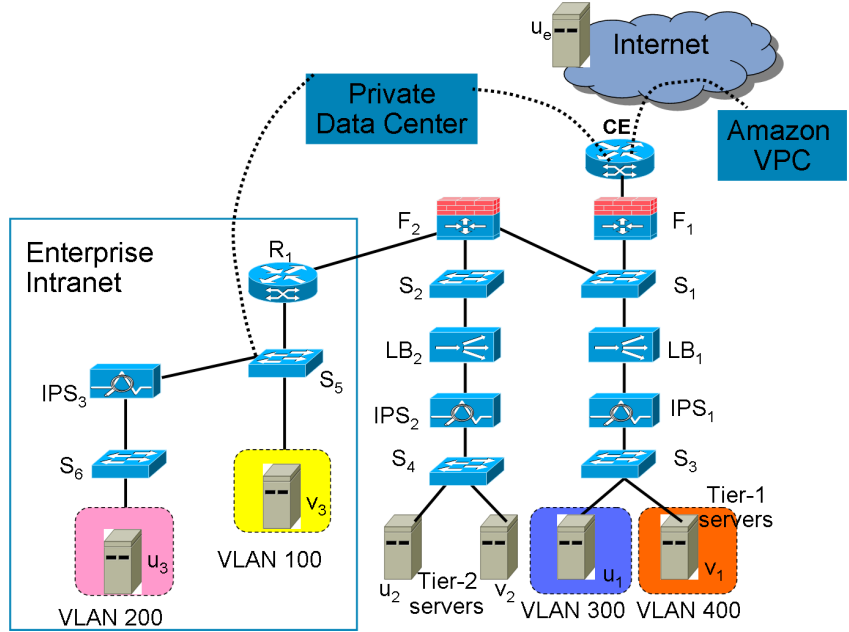

Figure 1: Motivating example.

Specifically, the tier-1 servers are the front ends of multiple network applications. The tier-1 servers of a given application are configured to belong to an IP subnet with private IP addresses. Each application is also assigned a public IP address to allow external access. Public IP addresses are assigned to the two load balancers represented by $LB_1$. A given application uses one load balancer as the primary and the other as the standby.

The tier-1 servers communicate with the tier-2 servers, which are located behind the two load balancers represented by $LB_2$. The tier-2 servers and $LB_2$ are configured with private IP addresses for security protection. $LB_1$ are configured with static routes to reach $LB_2$. The network border gateway *CE* has no knowledge about the routes to the tier-2 servers.

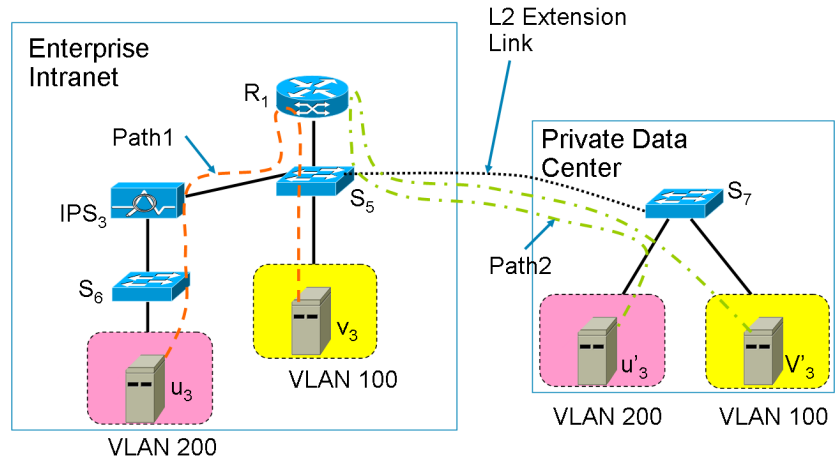

Figure 2: Problem with L2 extension.

Let us consider the possibility of relocating the tier-1 servers to a public cloud such as Amazon's EC2. One might consider this a trivial task. Specifically, after relocating the tier-1 servers to EC2, the operator simply updates $LB_1$ with the new IP addresses, if the IP addresses have to change. However, this simple solution can be broken in multiple aspects:

- *Violation of security policies:* The tier-1 servers are configured with multiple subnets, and the two boxes represented by $IPS_1$ monitor cross-subnet traffic. By simply relocating the tier-1 servers without relocating $IPS_1$, the solution bypasses the protection provided by $IPS_1$, violating the security policies of the organization.
- *Broken client TCP sessions:* Consider that an Internet client establishes a connection with a public IP address of $LB_1$. The load balancer directs the request from the client to one of the relocated tier-1 servers. The tier-1 server processes the request and sends back a reply, with the client's address as the destination and the server's address as the source. However, the client is expecting a reply with a source IP address of the load balancer, not the server. This breaks the client TCP session.
- *Disconnection from Tier-2 servers:* Recall that only $LB_1$ have routes to the tier-2 servers. Thus, when packets sent by the relocated tier-1 servers to tier-2 reach *CE* (the customer gateway), say via an Amazon VPC tunnel, *CE* will drop these packets because it does not know how to forward them.

In light of these issues, one may think they can be addressed by L2 exten-sions [2]. L2 extensions enable a LAN to be extended to a remote site. L2 extensions reduce network and application changes needed to support live server migration. Previous work has focused on transparency in terms of L2 connectivity [2, 8].

However, the remote data center may not support L2 extension. Furthermore, L2 extension still does not address policy homomorphism. In the preceding example, consider the case of extending both VLAN 100 and VLAN 200 into a remote data center. Because current public cloud infrastructure does not allow L2 extension, we focus on the case of extending to a private data center. Figure 2 zooms in on the left portion of Figure 1.

Assume that an L2 extension link is created between $S_5$ and $S_7$ and VLAN 100 and VLAN 200 are logically connected to $S_7$ in the remote data center. When a server $v_3$ in VLAN 100 communicates with $u_3$ in VLAN 200 in the enterprise network, the packet traverses: $v_1 \rightarrow S_1 \rightarrow R_1 \rightarrow S_1 \rightarrow IPS_1 \rightarrow S_2 \rightarrow u_1$. However, when $v'_3$ in VLAN 100 communicates with $u'_3$ in VLAN 200 in the private data center, it will not go through $IPS_3$; similarly the path from $v_3$ to $u'_3$ will not traverse $IPS_3$. Thus, L2 extension will not satisfy policy constraints automatically.

## 3 Mosaic Overview

The motivating example reveals potential issues facing the extension of an enterprise network into a remote data center. Mosaic is a systematic framework to address these issues. Mosaic consists of two major components: policy specification and network transformation.

**Policy specification:** To systematically investigate and solve the problems raised in the preceding section, we need to explicitly define the policies that an enterprise network intends to enforce so that one can validate any given solution. Policies capture the "invariants" that network extension should preserve. Since network extension alters an existing network topology (*e.g.*, by adding new nodes or relocating existing nodes), the *traversal* and *scope* of a packet (or frame if we talk about layer 2) can deviate from those in the original network. Thus, policy specification is crucial for policy enforcement, which will be discussed in Section 4.

**Network transformation:** Bounded by policy specification, network transformation computes the configuration at the remote data centers as well as at the local enterprise network. In addition to policies, multiple other factors, including objectives and constraints on application performance and migration costs, contribute to the complexity and effectiveness of network transformation.

The capabilities of network devices influence what transformation techniques may be used. In this paper, we do not assume the availability of futuristic mechanisms such as pswitches [5] and OpenFlow [6]. While these mechanisms can simplify our solutions, they have not been widely adopted so far. Instead, we only consider the traditional mechanisms that are readily available in today's enterprise networks. In Section 5, we will discuss the primitives and algorithmic framework of Mosaic.

## 4 Policy Specification

We start with the policy specification. We represent the topology of the original enterprise network $G$ using $V$, the set of nodes consisting of end hosts (servers, virtual machines), switches, routers and middleboxes; and $E$, the set of connections among network nodes.

An enterprise network operator defines policies $P$ on packets and frames, based on topology, as we have seen in the motivating example. Since we treat L3 packets and L2 frames uniformly in our framework, we use packet as a general term. For a given packet, policies specify additional information beyond what is already contained in the packet. Specifically, for a given packet $pkt_i$, policy `Policy`$_i$ consists of not only destination(s) `Destination`$_i$ but also two additional perspectives: waypoints `Waypoints`$_i$ and scope `Scope`$_i$.

By default, packets not associated with any policy are unwanted. These packets must be filtered before reaching their destinations. This default policy captures unreachability policies which are typically enforced by limiting route redistributions and specifying access control lists (ACLs) in routers.

**Waypoints:** The waypoints of a packet are the network nodes in addition to the destination(s) that should receive the packet. An enterprise may design its network such that a packet should pass through a particular set of network nodes. In the motivating example, we see that packets from the Internet should visit an intrusion prevention box before reaching a tier-1 server. As another example, an enterprise network may deploy a sniffer that is connected to the mirror port of a switch to receive a copy of a given packet for logging purpose. In this case, the sniffer also belongs to the waypoints of the given packet. Let `Waypoints`$_i$ be the waypoints of packet $pkt_i$.

Waypoints are specified by using the *ordering* and *occurrence* constraints. Ordering specifies if there are any constraints on the order to visit the nodes in the waypoints. For example, an enterprise network may require a packet to visit one middlebox before visiting another one. Occurrence specifies the number of times that a middlebox should be visited. For example, a packet may visit a middlebox only once, or none at all. We write `Waypoints`$_i$(`Order`$_i$, `Occurrence`$_i$) to emphasize that `Waypoints`$_i$ requires the ordering and occurrence constraints for $pkt_i$.

It is important to realize that we use network nodes in a generic sense when specifying waypoints. We can view each network node, in particular, a middlebox, as the member of a function class (*e.g.*, firewall, intrusion prevention, or sniffer) with a specific configuration state. Formally, we denote the function class of the middlebox node $v_j$ as `class(v_j)`; and its configuration state as `conf(v_j)`.

As an example, consider the network in Figure 1. The tier-1 and tier-2 firewalls have the same function class: `class(F_1) = class(F_2) = Firewall`. But their configuration states are different: the tier-1 firewall is in charge of the first line of defense and thus is configured to allow only HTTP traffic; the tier-2 firewall handles traffic from the tier-1 servers and intranet and thus may allow more protocols.

**Scope:** Destinations and waypoints capture the nodes that a packet *must* visit. However, a packet *may* reach other nodes in an enterprise network. For example, a modern switch may flood a given packet to a layer 2 domain if a forwarding entry is not present in its layer 2 FIB (forwarding information base); routers and switches along the path from the source to the destinations will see the packet (if unencrypted); due to routing changes, some routers not on the normal forwarding path may also see the packet. We associate a *scope* with each packet, which defines the security zone of the packet. The scope is the maximum set of nodes that a packet can reach. Let $\text{Scope}_i$ be the scope of *pkt$_i$*.

**Example Policies:** We now illustrate the preceding concepts using the example shown in Figure 1. Figure 3 specifies six policies for the network.

Policy `Policy_1` specifies that any HTTP request packet *pkt$_1$* to a tier-1 application server from an Internet client must traverse tier-1 firewall, tier1 load balancer (the tier-1 application's public accessible IP is configured at the load balancer $LB_1$). The packet's destination is changed to a tier-1 server $u_1$ by $LB_1$. We treat this as a new packet *pkt$_2$*. This packet with source $u_e$ and destination $u_1$ which originates from $L_1$ needs to traverse $IPS_1$. The scope of *pkt$_1$* $\text{Scope}_1 = \{\text{LB}_1, F_1, CE, S_1, u_e\}$. The scope of *pkt$_2$* $\text{Scope}_2 = \{\text{LB}_1, \text{IPS}_1, S_3, u_1\}$.

Policy `Policy_3` says that, any reply packet *pkt$_3$* from a tier-1 server to an Internet client must be sent to the load balancer first. It should be checked by $IPS_1$.

Policy `Policy_4` says that, for any packet *pkt$_4$* with source $LB_1$ originating from $u_1$, destined to an Internet client needs no further checks. $\text{Scope}_3 = \text{Scope}_2$ and $\text{Scope}_4 = \text{Scope}_1$.

Policy `Policy_5` states that a tier-1 server's packet *pkt$_5$* must traverse tier-2 firewall and load balancer $LB_2$. The scope $\text{Scope}_5 = \{u_1, u_2, F_2, \text{LB}_2, \text{IPS}_2, S_1, S_2, S_3, S_4, IPS_1, LB_1\}$.

Policy `Policy_6` states that cross-traffic between tier-1 servers in different subnet must be checked by $IPS_1$. The scope $\text{Scope}_6 = \{u_1, v_1, \text{IPS}_1, S_3\}$.

```
// 1. Internet client u_e to a tier-1 application
Policy_1 = ([u_e, L_1, *, 80, TCP], Scope_1, Waypoints_1({F_1LB_1},
           {σ|Ocurr(σ, F_1) = 1, Ocurr(σ, LB_1) = 1})
Policy_2 = ([u_e, u_1, *, 80, TCP], Scope_2, Waypoints_2({IPS_1},
           {σ|Ocurr(σ, IPS_1) > 0})

// 2. Tier-1 application server u_1's reply to Internet client u_e
Policy_3 = ([u_1, u_e, 80, *, TCP], Scope_3,
           {σ|Ocurr(σ, LB_1) = 1, Ocurr(σ, IPS_1) > 0})
Policy_4 = ([u_1, u_e, 80, *, TCP], Scope_4, Waypoints_2({}, {}))

// 3. Tier-1 application server u_1 communicates with tier-2 server u_2
Policy_5 = ([u_1, u_2, *, *, TCP], Scope_5, {F_2LB_2IPS_2},
           {σ|Ocurr(σ, F_2) = 1, Ocurr(σ, LB_2) = 1, Ocurr(σ, IPS_2) > 0})

// 4. Tier-1 application server u_1 in subnet 1 communicates with tier-1
//    application server v_1 in subnet 2
Policy_6 = ([u_1, v_1, *, *, TCP], Scope_6, Waypoints_6({IPS_2},
           {σ|Ocurr(σ, IPS_2) > 0})
```

Figure 3: Policies for enterprise network in Figure 1.

# 5  Network Transformation

We consider network transformation algorithms that take as input an original network, its policy specification, the set $U$ of servers to be extended or relocated to remote data centers, reconfiguration constraints at the local and remote data centers, the cost model of network equipment and traffic, and performance constraints on applications. The set $U$ can be manually given or computed by another algorithm.

The outputs of network transformation include:

- the connecitvity from local data centers to remote data centers;
- the restructuring of the local data center, including addition and deletetion of nodes, as well as reconfiguration of existing nodes;
- the configuration of the remote data centers.

Note that the capabilities supported at the remote data center can place substantial constraints on the outputs of the network transformation algorithm. Consider Amazon's VPC as an example remote data center. VPC makes public cloud resources appear the same as internal enterprise resources. However, VPC imposes specific constraints on the connections from the enterprise network. First, VPC specifies L3 connectivity. Second, inside VPC, the enterprise can construct only a logical star topology connecting multiple subnets. On the other hand, a private data center, for instance, a new data center owned by the same enterprise, may allow more flexibility. In this case, the remote data center may allow both L2 and L3 connectivity from the local enterprise network to the remote data center. Also, the enterprise can have flexibility in constructing a topology and placing policy devices inside the remote data center.

To be concrete, we present a two-stage transformation algorithm.

**Stage 1:** The algorithm computes, for each policy, whether to enforce it at the local data center or the remote data center. The computation is based on the constraints on application performance (*e.g.*, delay constraints), enterprise costs (*e.g.*, cross data-center traffic and equipment replication cost), and the availability of policy classes at the local and remote data centers.

**Stage 2:** The algorithm constructs detailed configurations at the local and remote data centers.

Instead of going over all steps of the complete algorithm, we present three key primitives used at Stage 2:

- Mosaic proxy: this primitive allows enforcement of a policy at the local data center. The primitive is driven by the principle of least-disruption and greatest re-use. It enforces a policy by traversing the original policy boxes in the local data center. Thus, it has low equipmemnt cost. It avoids policy ommisions such as those discussed at the end of Section 2.

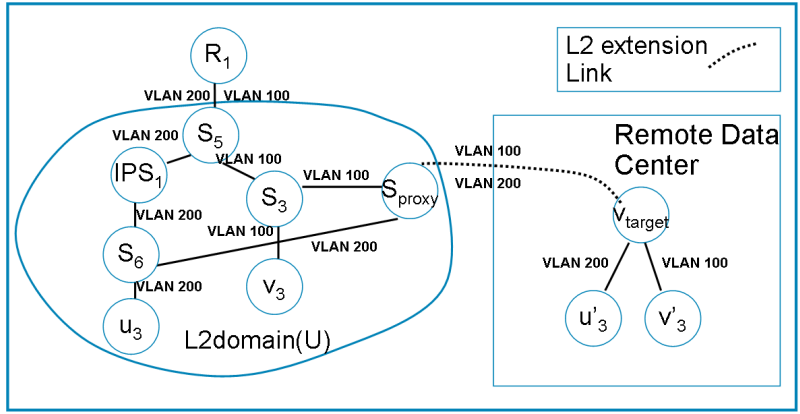

Figure 4: Mosaic proxy resolves policy violation of Fig 2.

- Mosaic mirror: this primitive enforces policies at the remote data center by replicating a minimal set of policy boxes in the remote data center. The replicated set is achieved by computing an edge-cut-set surrounding relocated nodes to ensure robust policy enforcement even in the presence of failures. Enforcing policies at the remote data center reduces latency, in particular, for traffic among relocated servers.

- Mosaic policy relocaton: this primitive optimizes specific classes of policies (*e.g.*, firewall) by relocating them from one device to another existing network device (*e.g.*, as a different firewall context) to enforce policy without introducing any new devices.

In this paper, we present more details only for the Mosaic proxy primitive, which forces packets to traverse the original policy at the local data center.

Figure 4 illustrates the introducion of a Mosaic proxy to fix the problem of policy violation using standard L2 extension, as shown in Figure 2. Let $v_{target}$ denote the entrance to the remote data center. Mosaic proxy introduces a switch $S_{proxy}$ with L2 connectivity to $v_{target}$. Since $u_3$ and $v_3$ will migrate to the remote data center, Mosaic proxy connects their corresponding switches $S_5$ and $S_6$ to $S_{proxy}$ with VLAN configurations shown in the figure.

Now, consider the policy that communications between $u'_3$ and $v'_3$ be checked by $IPS_1$ in the network after migration. Specifically, since $v'_3$ and $u'_3$ are in different subnets, a packet from one to the other will be routed to $R_1$ in VLAN 100. In particular, the path from $v'_3$ to $R_1$ is $v'_3 \to v_{target} \to S_{proxy} \to S_3 \to S_5 \to R_1$. The path from $R_1$ to $u'_3$ in VLAN 200 is $S_5 \to IPS_1 \to S_6 \to S_{proxy} \to v_{target} \to u'_3$.

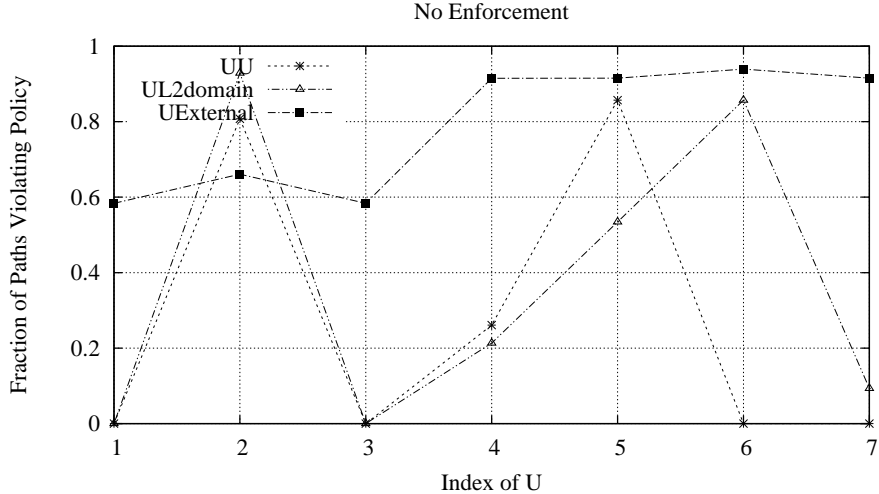

Figure 5: Fraction of paths with policy violations without Mosaic.

Thus, any packet from $v'_3$ to $u'_3$ traverses the policy box $IPS_1$, satisfying the policy requirement.

Note that $S_{proxy}$ does not have to be a new device. It can be any L2 switch that can connect to $V_{target}$. The links connecting $S_5, S_6$ to $S_{proxy}$ can be implemented logically using private VLANs, in order to make sure that switches $S_5, S_6$ can talk to only $S_{proxy}$, not between themselves.

## 6 Evaluation

We conduct preliminary evaluation on the effectiveness of Mosaic. Specifically, we obtain router, middlebox and switch configuration files of a campus network with more than 50 routers and more than 1000 switches. We extract route distribution graph, and L3 topology using a tool in [1]. We then insert the L2 topology into L3 topology due to the fact that switch configurations are not adequate. We infer the middlebox traversal policy based on the topology properties and route distribution graph. We examine the possible paths between two endpoints (represented as two subnets or two VLANs). From the path, we determine the middleboxes traversed and store this sequence as the way-points for this particular path. For scope, if both endpoints are in the same VLAN, then the scope is all nodes in the broadcast domain. If they are not in the same VLAN, we use all reachable nodes (based on route distribution graph and ACLs) in the security zone as the scope.

Figure 5 shows the results of no policy enforcement extension, when we pick a set $U$ of servers to relocate. The x-axis is the size of the set of servers to relocate. Specifically, the algorithm operates in the following way: Routes in the remote data center are advertised through BGP and are redistributed to the enterprise network. As such, the design is the most efficient solution in terms of performance. However, it is not a viable option to enforce policy requirements due to its volatile tendency to violate policy. We break the communication into three types: `UU` sessions (among relocated servers), `UL2domain` (between relocated and those remaining at the same L2 domain), and `UExternal`(between relocated and those outside the campus network). We see that there can be significant policy violations. The `UExternal` policy violations can occur on as many as 80% of the paths due to traversal of several security zones. Conversely, Mosaic sustains no policy violations.

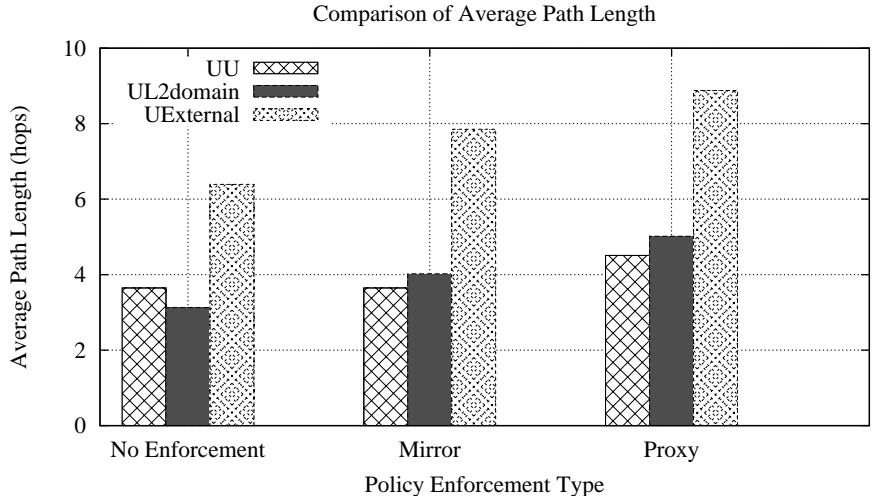

Figure 6: A comparison of the average path length for all communication sessions.

Although the no-policy-enforcement approach is not feasible in practice, it is a baseline comparison for measuring the cost of enforcing policy. In the next experiment, we evaluate the cost of enforcing policies by considering the average path length for communication between points in relocated $U$ and other endpoints in the network. The path length is defined as the number of network devices a packet must traverse from source to destination (*i.e.*, network hops).

Figure 6 shows the results. The distances shown in these results only include the enterprise portion and count both L2 and L3 hops. A tunnel hop is counted by the number of hops it traverses. It is important to recall that Mosaic-Mirror enforces policies remotely, whereas Mosaic-Proxy enforces locally. We observe that the price of enforcing policies measured by hop counts may not be significant, shown by the non-significant increase of hop counts compared with no policy enforcement.

# 7 Conclusion and future work

Network extension and migration are now a major challenge for large-scale networks, attracting industrial attention (*e.g.*, [2, 4, 3, 8]). Ad-hoc methods of network extension and migration can result in serious policy violations. In this paper, we present the first framework for network policy specification. Furthermore, we evaluate the feasibility of policy homomorphic network extension and migration to remote data centers.

# References

[1] T. Benson, A. Akella, and D. Maltz. Mining policies from enterprise network configuration. In *IMC'09*, 2009.

[2] Cisco Inc. Data center interconnect: Layer 2 extension between remote data centers.

[3] Cohesive Flexible Technologies Corp. VPN-Cubed: customer controlled security for the cloud.

[4] Computer World. IBM, Juniper join in cloud strategy: Technology would reallocate computing resources between private, public cloud, Feb. 2009.

[5] D. Joseph, A. Tavakoli, and I. Stoica. A policy-aware switching layer for data centers. *CCR*, 2008.

[6] Openflow. The openflow switch specification.

[7] SearchDataCenter and Viridity. An expert guide to data center trends, energy regulations, and more, 2010.

[8] T. Wood, A. Gerber, K. K. Ramakrishnan, and J. Van Der Merwe. The case for enterprise-ready virtual private clouds. In *USENIX HotCloud*, June 2009.